\documentclass[reprint,prb,superscriptaddress]{revtex4-1}

\usepackage{amsmath}
\usepackage{amsfonts}
\usepackage{amssymb}
\usepackage{graphicx}
\usepackage{moreverb}
\usepackage{epsf}
\usepackage{color}
\usepackage{multirow}
\usepackage{subfigure}

\newcommand{\DZP}{$\mathrm{d}\zeta+\mathrm{p}$}
\newcommand{\QZDP}{$\mathrm{q}\zeta+\mathrm{dp}$}

\begin{document}

\title{Performance analysis of electronic structure codes on HPC systems:\\ A case study of SIESTA}
\author{Fabiano~Corsetti}
\email[E-mail: ]{f.corsetti@nanogune.eu}
\affiliation{CIC nanoGUNE, 20018 Donostia-San Sebasti\'{a}n, Spain}
\date{\today}

\begin{abstract}
We report on scaling and timing tests of the SIESTA electronic structure code for {\em ab initio} molecular dynamics simulations using density-functional theory. The tests are performed on six large-scale supercomputers belonging to the PRACE Tier-0 network with four different architectures: Cray XE6, IBM BlueGene/Q, BullX, and IBM iDataPlex. We employ a systematic strategy for simultaneously testing weak and strong scaling, and propose a measure which is independent of the range of number of cores on which the tests are performed to quantify strong scaling efficiency as a function of simulation size. We find an increase in efficiency with simulation size for all machines, with a qualitatively different curve depending on the supercomputer topology, and discuss the connection of this functional form with weak scaling behaviour. We also analyze the absolute timings obtained in our tests, showing the range of system sizes and cores favourable for different machines. Our results can be employed as a guide both for running SIESTA on parallel architectures, and for executing similar scaling tests of other electronic structure codes.
\end{abstract}

\maketitle

\section{Introduction}

The use of first principles atomistic simulations with density-functional theory~\cite{hk,ks} (DFT) has grown from a cottage industry in the early 1990s to a routine and integral part of many contemporary scientific disciplines, at the meeting point between condensed matter physics, physical chemistry, and the new range of nanosciences~\cite{intro1,intro2}. Potential practitioners have a large number of ready-made codes to choose from (see, e.g., Refs.~[\onlinecite{Kresse1996a,Delley2000,Soler2002,castep,VandeVondele2005103,Skylaris2005,Gygi2008,BigDFT,QE-2009,Blum20092175,abinit-generic,Enkovaara2010}]), which distinguish themselves in their licensing models, the range of features they offer, the specifics of the technical implementation, and, generally, where they lie on the (computational) cost--accuracy curve.

An important consideration for all modern DFT codes is their parallel scalability on high-performance computer (HPC) architectures, that open up the possibility of simulating very large physical systems entirely {\em ab initio}. Consequently, a substantial effort has gone into the development and optimization of many of these codes for the specific purpose of running on massively parallel systems~\cite{B600182N,Bottin2008329,Gygi2008,Genovese2009,onetep-nick,Sanz-Navarro2010,Bowler2010,Auckenthaler2011783,Maniopoulou20121696,VandeVondele2012,JCC:JCC23096,Varini20131827,GPAW-GPU}. Articles describing such developments typically illustrate the scaling performance of the code with an example of strong scaling~\cite{VandeVondele2005103,Skylaris2005,Gygi2008,BigDFT,QE-2009,Blum20092175,abinit-generic,Enkovaara2010,B600182N,Bottin2008329,Genovese2009,onetep-nick,Sanz-Navarro2010,Bowler2010,Auckenthaler2011783,Maniopoulou20121696,VandeVondele2012,JCC:JCC23096,Varini20131827}, i.e., the wall time speedup obtained for a simulation of fixed size over a range of number of cores. Less frequently, weak scaling performance (i.e., an increase of the problem size proportionally with the number of cores) is also shown~\cite{Bowler2010,VandeVondele2012,GPAW-GPU}.

The use of a strong scaling example can be an effective way of giving a qualitative idea of the parallel efficiency of the code and the scale of problems which can realistically be solved with it. However, there are a number of issues in using the information as it is usually presented for extracting, even approximately, a generalized, quantitative measure of performance, such as could be used to attempt a comparison between codes.

Firstly, the range of cores over which this strong scaling is investigated is not fixed (as must be the case, since time and memory requirements restrict the lower bound, and computational resources the upper bound). The significance of the demonstrated speedup depends crucially on the lower bound of this range; furthermore, the dependence is non-trivial. If we assume a constant rate of loss of efficiency as the parallelization is increased, a speedup of 3.9 when going from 8 to 32 cores should be better than a speedup of 3.8 when going from 2048 to 8192 cores; however, this is obviously not the case, as it is clear from experience that the actual rate increases significantly with the number of cores. Closer comparisons are even harder to judge: is a speedup of 3.8 between 512 and 2048 cores better or worse than a speedup of 3.7 between 1024 and 4096 cores? There is effectively no way to answer this question without making an assumption about how to model parallel performance in general. A well-known and popular, albeit extremely idealized, way to do is by Amdahl's law~\cite{amdahl}, that describes the overall speedup in terms of the parallelizable fraction of the code $P$. To the best of our knowledge, only one published strong scaling test for a DFT code~\cite{onetep-nick} has reported on a fitted value for $P$.

Secondly, there is no standard physical system on which to test strong scaling. From the point of view of the material itself, this is somewhat understandable, as different codes specialize in different areas of modelling; a more fundamental problem, however, is that strong scaling efficiency changes with system size for a given material. Although some studies report system size dependent results~\cite{VandeVondele2005103,BigDFT}, this is generally not the case. How, then, to compare between, e.g., a strong scaling test on a 1532-atom carbon nanotube between 2048 and 32768 cores~\cite{Varini20131827}, and one on a 1003-atom polyalanine peptide between 512 and 65536 cores~\cite{Auckenthaler2011783}?

\begin{table*} \label{table:specs}
\begin{tabular*}{0.93\textwidth}{lccccccccc}
\hline
\hline
            &                &          &                   & Proc.       & Tot.   & Tot.  & Cores/ & Cores/ & Mem./     \\
System      & Architecture   & Topology & Processor type    & speed (GHz) & cores  & nodes & node   & proc.  & core (GB) \\
\hline                                                                                                                                                         
Hermit      & Cray XE6       & 3D torus & AMD Opteron       &         2.3 & 113664 &  3552 &     32 &     16 &       2/4 \\
JUQUEEN     & IBM BlueGene/Q & 5D torus & IBM PowerPC A2    &         1.6 & 458752 & 28672 &     16 &      8 &         1 \\
FERMI       & IBM BlueGene/Q & 5D torus & IBM PowerPC A2    &         1.6 & 163840 & 10240 &     16 &      8 &         1 \\
Curie       & BullX          & Fat tree & Intel SandyBridge &         2.7 &  80640 &  5040 &     16 &      8 &         4 \\
SuperMUC    & IBM iDataPlex  & Fat tree & Intel SandyBridge &         2.7 & 147456 &  9216 &     16 &      8 &         2 \\
MareNostrum & IBM iDataPlex  & Fat tree & Intel SandyBridge &         2.6 &  48384 &  3024 &     16 &      8 &         2 \\
\hline
\hline
\end{tabular*}
\caption{Specifications for the six PRACE Tier-0 systems. We note that some systems include secondary types of nodes with different specifications; these are not listed here, and are not used for our tests.}
\end{table*}

In this paper, we discuss these issues while reporting on tests of the parallel scaling performance of SIESTA~\cite{Soler2002}, a well-established DFT code based on norm-conserving pseudopotentials~\cite{Troullier1991}, a basis of finite-range numerical atomic orbitals (NAOs), and an auxiliary real-space grid for representing the electronic density. The tests are performed on six supercomputers (Table~\ref{table:specs}), currently forming the network of Tier-0 systems of the Partnership for Advanced Computing in Europe~\cite{PRACE} (PRACE). Our aims, therefore, are twofold:
\begin{itemize}
\item to give the most up-to-date, comprehensive and reliable results of the timing and scaling of SIESTA on modern HPC systems, so as to allow users of the code to calculate  realistic timing estimates over a wide range of number of cores, and therefore plan how to make the best use of their computational resources;
\item to propose a simple framework in which to analyze parallel scaling results for all electronic structure codes, arguing in particular for the use of Amdahl's law to quantify strong scaling performance, and for the importance of investigating and reporting this measure as a function of system size.
\end{itemize}

\section{Computational methods}

Our scaling tests are performed on snapshots of liquid water in cubic boxes with periodic boundary conditions. This is the same system used previously for parallel benchmarking of the Quickstep~\cite{VandeVondele2005103} (CP2K) code; as noted by its authors, liquid water is ideal for this purpose, since boxes of any arbitrary number of molecules can be created while maintaining the same density and cell shape. Furthermore, the lack of crystalline symmetry and the 3D periodicity of the material ensure a sufficiently challenging task that we expect to give a fair idea of worst-case timings for most typical uses of the code, while the presence of a band gap ensures that we do not have to worry about convergence issues arising during the tests.

\begin{figure}
\begin{center}
\includegraphics{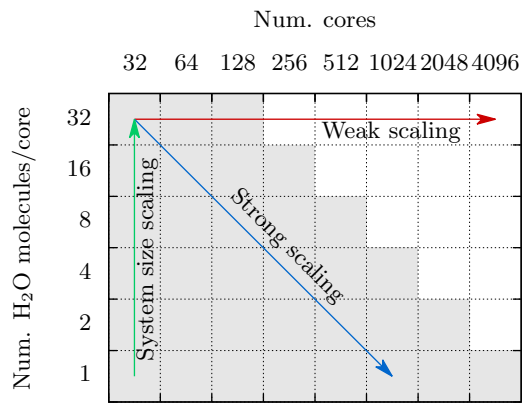}
\end{center}
\caption{Three types of scaling that can be investigated by systematically varying the number of molecules per core and the number of cores. The shaded cells show the suggested set of tests to perform on a typical HPC system.}
\label{fig:scaling}
\end{figure}

We simultaneously test weak and strong scaling (again similarly to Ref.~[\onlinecite{VandeVondele2005103}]), by varying both the number of cores, from 32 to 4096 ($N_c = 2^n, 5 \le n \le 12$), and the number of water molecules per core, from 1 to 32 ($N_m/N_c = 2^n, 0 \le n \le 5$). The resulting suite of tests is shown in Fig.~\ref{fig:scaling}. The maximum system size tested is of 4096 water molecules (12288 atoms) for all values of $N_m/N_c$, except for one test of 8192 molecules (24576 atoms) on 8192 cores. We note, however, that due to the limited computational time available on each machine, not all tests are run on all machines. Weak scaling corresponds to moving perpendicular to the $N_m/N_c$ axis, while strong scaling corresponds to moving diagonally. Instead, system size scaling (parallel to the $N_m/N_c$ axis) does not explicitly test parallelization, although it is affected by it, as we shall discuss. The snapshots for all system sizes are extracted from classical molecular dynamics (MD) runs using the TIP4P force field~\cite{Jorgensen1983b} in the GROMACS~\cite{gromacs} code, equilibrated to 300~K; the cell shape and volume are kept fixed at the experimental equilibrium density~\cite{water_density} (1.00~g/cm$^3$).

The tests are performed at the $\Gamma$ point only (multiple k points being almost embarrassingly parallel), using the semi-local PBE~\cite{pbe} functional for exchange and correlation (xc), a 150~Ry cutoff energy for the real-space auxiliary grid, and, unless otherwise stated, a double-$\zeta$ polarized basis~\cite{water_emiliomarivi} (\DZP), corresponding to 23 NAOs per water molecule; the fraction of occupied eigenstates is $4/23$ ($\sim$17\%). All system sizes employ 13 self-consistent field (SCF) iterations to reach convergence.

We use the most recent development version of the code ({\tt siesta-trunk-438}), available on the SIESTA website~\cite{SIESTA-website}. The tests are run with the code's default options for diagonalization, employing routines from the ScaLAPACK~\cite{slug} library: the problem is first transformed from generalized to standard form by Cholesky factorization with the {\tt pdpotrf} and {\tt pdsygst} routines, and then the diagonalization itself is performed with the {\tt pdsyevd} divide-and-conquer routine; finally, the back transform is performed with the {\tt pdtrsm} routine. A 2D block-cyclic data distribution of the matrices is used, with the matrix dimension being an exact multiple of the block size in all cases (tests show the ideal block size to be equal to the number of orbitals per molecule).

We choose the standard solver for our tests, as this is currently by far the most widely used by the SIESTA community; however, we note that several new alternatives are being developed and tested: (i) a solver based on the orbital minimization method (OMM), which has already been demonstrated to exhibit better parallel scaling than explicit diagonalization up to 64 cores~\cite{OMM} (available in the development version of the code), (ii) two new solvers based on ScaLAPACK, the MRRR algorithm~\cite{Antonelli:CSD-05-1399} and the ELPA library~\cite{Auckenthaler2011783} (not yet released), and (iii) a solver based on the pole expansion and selected inversion method~\cite{Lin2013}, specifically designed for massively parallel architectures (not yet released). Finally, the original linear-scaling DFT method implemented in SIESTA is also in the process of being redesigned; in its current implementation it does not scale well on large clusters.

The code was compiled on each of the six machines listed in Table~\ref{table:specs} using the native Fortran compiler and optimized linear algebra and communication libraries provided by the system administrators. The Intel compiler and MKL library are used for Intel-based machines (IBM iDataPlex and BullX architectures), the Cray compiler and ACML library for the AMD-based machine (Cray XE6 architecture), and the IBM XL compiler and ESSL library for the IBM PowerPC-based machines (IBM BlueGene/Q architecture). The MPI-2 libraries used are as follows: IBM MPI for SuperMUC, Open MPI for MareNostrum, BullX MPI for Curie, and MPICH2 for Hermit, JUQUEEN and FERMI.

\begin{figure}
\begin{center}
\includegraphics{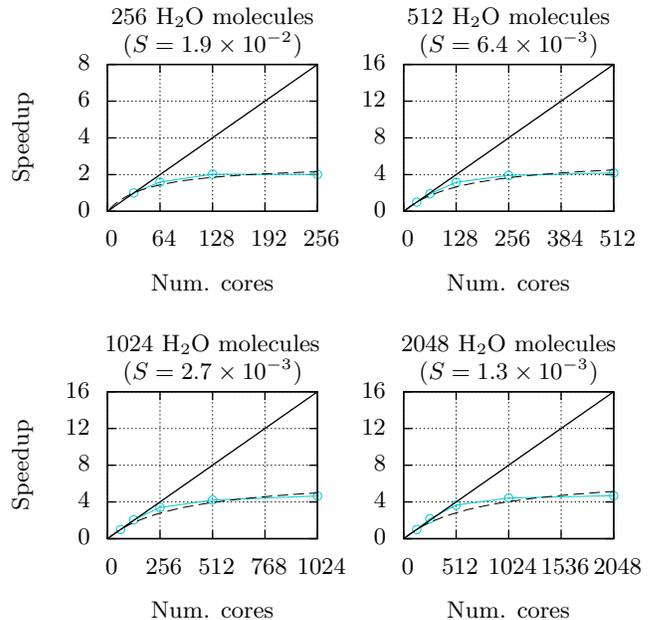}
\end{center}
\caption{Strong scaling on SuperMUC for four different system sizes. The full black lines gives the ideal scaling relative to the smallest system size. The fit to Amdahl's law is shown by the dashed black line, and the corrsponding $S$ value is given above the plot.}
\label{fig:SuperMUC}
\end{figure}

\section{Results and discussion}

\subsection{Strong scaling}

\begin{figure}
\begin{center}
\includegraphics{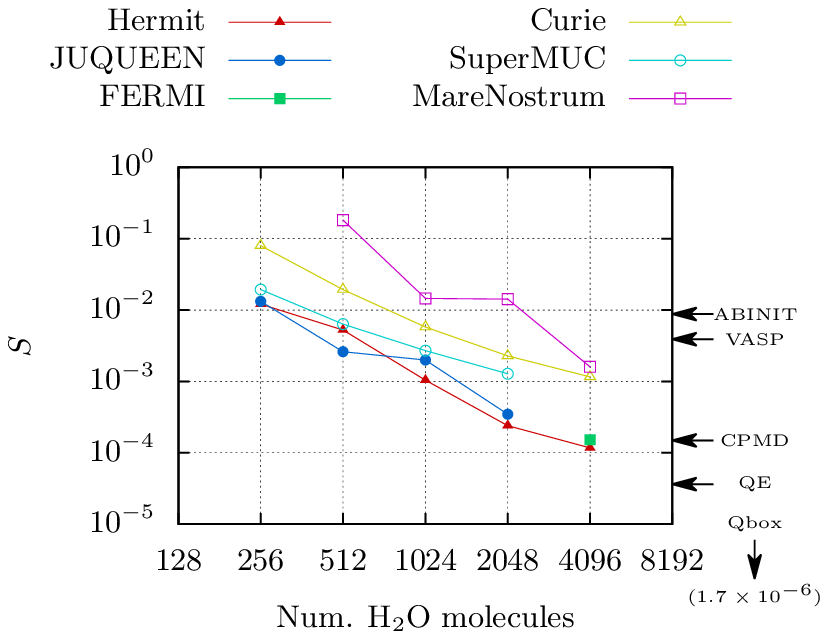}
\includegraphics{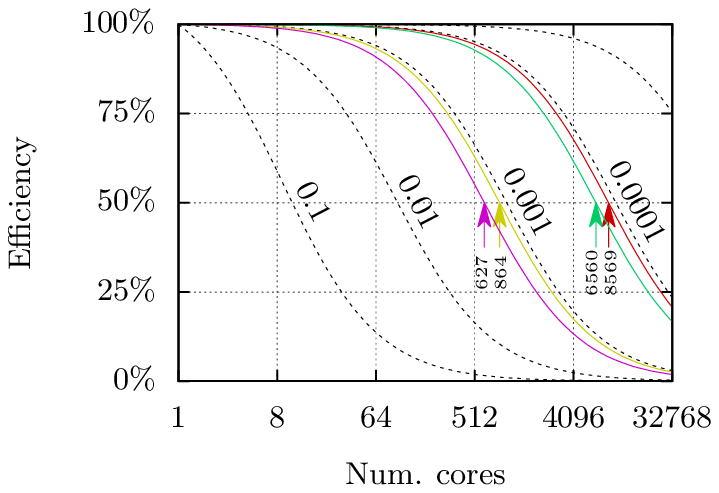}
\end{center}
\caption{Strong scaling and efficiency. Top panel: $S$ value as a function of system size fitted to strong scaling data obtained with SIESTA on the six machines; also included are values calculated with other DFT codes for a single system size on IBM BlueGene architectures (ABINIT: 108 atoms, 1188 electrons, 3D system, 4 k points~\cite{FERMI-benchmarks}; VASP: 87 atoms, 822 electrons, 2D system, 14 k points~\cite{FERMI-benchmarks}; CPMD: 284 atoms, 1192 electrons, 3D system, k-point sampling unspecified~\cite{FERMI-benchmarks}; QE: 1532 atoms, 5232 electrons, 1D system, $\Gamma$ point~\cite{FERMI-benchmarks,Varini20131827}; Qbox: 1000 atoms, 12000 electrons, 3D system, $\Gamma$ point~\cite{Gygi2008}). Bottom panel: relationship between $S$ and core hour efficiency as a function of the number of cores, for four different values of $S$ given by the black dashed lines, and the fitted values of $S$ obtained with SIESTA on four different machines for a system of 4096 water molecules; the number of cores at which the efficiency is equal to 50\% is labelled in each case.}
\label{fig:law}
\end{figure}

As previously mentioned, Amdahl's law provides a simple model of strong scaling. It states that
\begin{equation} \label{eq:law1}
\mathbb{S}_1 \left ( N_c; S \right )=\frac{t_1}{t_{N_c}}=\frac{1}{S+\frac{1-S}{N_c}},
\end{equation}
where $\mathbb{S}_1$ is the speedup obtained on $N_c$ cores with respect to a serial run, $t_1$ and $t_{N_c}$ are the total execution times in serial and on $N_c$ cores, respectively, and $S=1-P$ is the fraction of the code that is not parallelizable (we prefer using $S$ instead of the more usual $P$, as the former tends to zero in the limit of ideal scaling). Since it is usually not possible in practice to measure $t_1$ for large systems, it is useful to define the speedup with respect to a baseline number of cores $b$ instead:
\begin{equation} \label{eq:law}
\mathbb{S}_b \left ( N_c; S \right )=\frac{t_b}{t_{N_c}}=\left ( \frac{N_c}{b} \right ) \frac{S \left ( b-1\right )+1}{S \left ( N_c-1\right )+1}.
\end{equation}
Using this equation, we can fit our strong scaling data over any arbitrary range of number of cores, and obtain a single value $S$ that is in principle independent of this range, and which therefore defines the efficiency of the code for any value of $N_c$ as $1/ \left ( 1+S \left ( N_c-1 \right ) \right )$ (see bottom panel, Fig.~\ref{fig:law}). The efficiency is invariantly 100\% for a serial run, and decreases to zero as $N_c \to \infty$, since the execution time tends to a finite minimum value $t_1 S$.

It is important to note that the conventional interpretation for $S$ and $P$ is necessarily an over-simplification, and should not be taken too literally; nevertheless, Amdahl's law qualitatively reproduces some universal features of strong scaling, and is generally found to provide a good fit to real data. However, such a basic one-parameter model can only describe an average scaling trend, ignoring any system dependent effects that might favour particular values of $N_c$, e.g., differences in load balancing. Using a homogeneous, scalable system such as liquid water and a regular grid of tests as shown in Fig.~\ref{fig:scaling} can be effective in minimizing these variations, and therefore help to extract clearer general trends.

Using our timing tests for SIESTA on the six different machines, we can analyze the strong scaling of the code for system sizes ranging from 64 to 4096 water molecules; however, we restrict our fitting of $S$ to systems with at least four data points ($\ge 256$ molecules). As a representative example, Fig.~\ref{fig:SuperMUC} shows the speedup obtained on SuperMUC (IBM iDataPlex architecture) for four different system sizes, together with the curve fitted from Eq.~\ref{eq:law}. The resulting $S$ values are robust to fitting over different ranges (within an order of magnitude), as is the trend of decreasing $S$ with increasing system size. It is worth noting that this example clearly illustrates the difficulty in comparing between scaling tests using different ranges of number of cores: despite the steady increase in efficiency revelead by the $S$ values, the speedups shown in the plots appear extremely similar due to the different baselines used.

The top panel of Fig.~\ref{fig:law} summarizes the strong scaling results obtained for all six machines: the fitted value of $S$ is given as a function of system size (i.e., the number of water molecules), for systems between 256 and 4096 molecules. Tests for smaller systems (32, 64, and 128 molecules) and larger ones (8192 molecules) are not represented, as there are insufficient data points for a reliable fit.

For all HPC systems, $S$ is observed to decrease with the size of the physical system being simulated. This should not be surprising, as it is reasonable to expect efficiency to be related to the number of matrix elements/core (which in turn determines the ratio of intracore to intercore operations), and, hence, that the larger the system being simulated, the larger the number of cores on which the calculation can be performed before the efficiency drops below a given threshold. However, the detailed form of this decrease depends on many factors related to the nature of the operations being performed and the computational architecture, and is therefore strongly dependent on the code and the HPC system used.

We can see some interesting distinctions in the $S \left ( N_m \right )$ curves for the six machines. There is a very close agreement between the three machines implementing torus topologies (Hermit, JUQUEEN, FERMI), despite Hermit being quite distinct from JUQUEEN and FERMI in most other respects, e.g., architecture type (Cray XE6 for the former, IBM BlueGene/Q for the latter) torus dimension, processor type and speed, number of cores per node and amount of memory per core. Instead, the three machines implementing fat tree topologies (Curie, SuperMUC, MareNostrum), even though they do not exhibit the same level of agreement amongst each other, give consistently higher $S$ values than the torus machines.

Furthermore, despite the limited data available, our results suggest a qualitatively different form of the decrease of $S$ with $N_m$ for machines with torus and fat tree topologies. The former show an approximately linear decrease with slope $B$ on the log--log scale ($S \propto N_m^{-B}$), while the latter exhibit a slowing down of the rate of decrease. This is confirmed by fitting the data for each machine to a quadratic polynomial on the log--log scale; we find that the quadratic coefficient, positive in all cases, is an order of magnitude ($\sim$3--15 times) smaller for the machines with torus topologies compared to those with fat tree topologies.

Using the $S$ value as a measure of strong scaling, we can attempt a quantitative comparison between SIESTA and other DFT codes; this is given alongside our results in the top panel of Fig.~\ref{fig:law}. The fits are performed using publicly available scaling test data for the codes, published on the website~\cite{FERMI-benchmarks} of the FERMI IBM BlueGene/Q machine, which we also use for our tests of SIESTA; the same data for the Quantum ESPRESSO (QE) code has also been published in an article describing development work on the code~\cite{Varini20131827}. The only exception is the Qbox code, for which we used previously published tests~\cite{Gygi2008} performed on an IBM BlueGene/L machine. Where possible, we select tests performed $\Gamma$-only. All codes considered employ a plane-wave basis, in contrast to SIESTA's much smaller NAO basis.

It is important to stress that this comparison serves mainly to highlight the inadequacy of the available data; indeed, the change in $S$ over more than three orders of magnitude for SIESTA at different system sizes is similar to the range spanned by the results obtained for the other codes, each available at a single system size. Both the system size and type vary greatly between codes, from an 87-atom 2D system for VASP to a 1532-atom 1D system for QE. Other important factors (k-point sampling, xc functional, basis accuracy, code optimization) are also not controlled for.

Nevertheless, Qbox stands out from all other codes for the impressive strong scaling performance demonstrated, with an $S$ value more than an order of magnitude lower than that obtained by its closest competitor, QE, despite using a smaller system size (1000 atoms). Indeed, Qbox has been developed not only for massively parallel calculations in general, but specifically for running on IBM BlueGene architectures~\cite{Gygi2006,Gygi2008}; based on these results, it is the only DFT code to have demonstrated the potential to make efficient use ($>$50\%) of the {\em entirety} of a large BlueGene machine such as JUQUEEN or FERMI for a single $\Gamma$-point calculation.

\subsection{System size scaling and absolute timings}

\begin{figure}
\begin{center}
\includegraphics{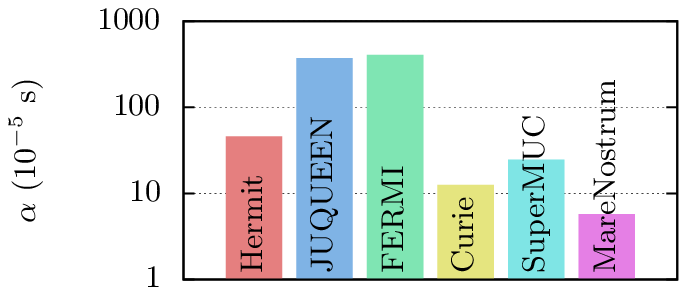}
\includegraphics{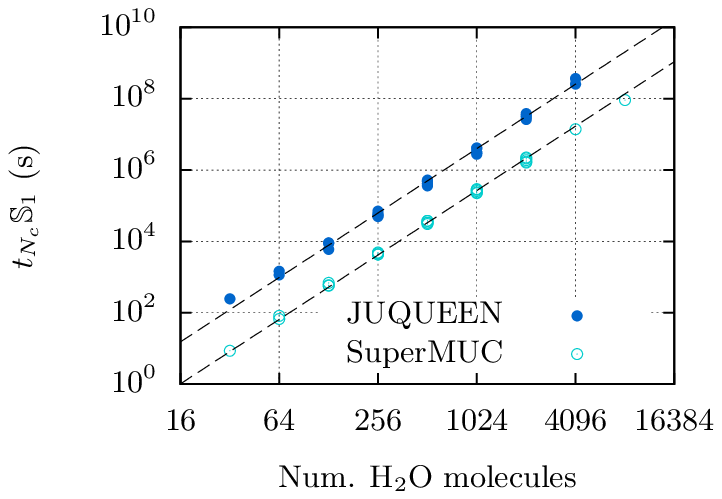}
\end{center}
\caption{Absolute timings on the six machines. Top panel: prefactor $\alpha$ for the cubic scaling with system size of the execution time in serial for the self-consistent calculation of the liquid water system (13 SCF iterations). Bottom panel: two examples of the fitting of $\alpha$ to absolute timing data, extrapolated for all number of cores to serial timings using Amdahl's law and a fitted analytical expression of the strong scaling performance as a function of system size.}
\label{fig:alpha}
\end{figure}

\begin{figure}
\begin{center}
\includegraphics{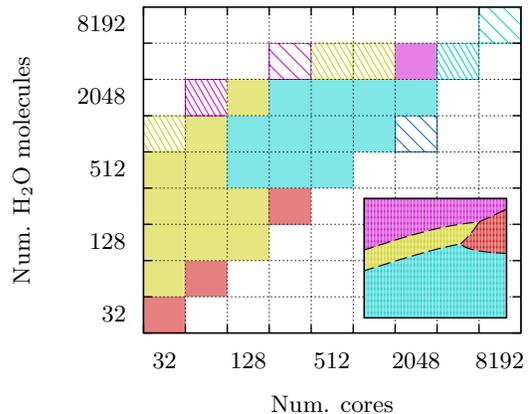}
\end{center}
\caption{Phase diagram of supercomputers. The machine with the lowest execution time is shown for a given system size and number of cores. The colours used are the same as those shown in the top panel of Fig.~\ref{fig:alpha}. Boxes with dashed lines indicate that the data for one or more machines is not available; sparse dashed lines indicate that only one machine was run with these parameters. The inset shows the idealized diagram over the same range, using the timing estimates given by the fits of $S \left ( N_m \right )$ and $\alpha$.}
\label{fig:phase}
\end{figure}

Strong scaling is purely a test of parallel scalability, for which the code is, by definition, taken to be 100\% efficient when run in serial. The results presented so far, therefore, contain no information about absolute timings. Although it is convenient to separate these two aspects of the code's performance, we should remember that the execution time is the {\em only} factor of importance to the end user. Therefore, strong scaling data on its own can sometimes be misleading, as a code that is very fast in serial but which exhibits poor strong scaling might nevertheless achieve a lower execution time on a medium-sized cluster than one that is very slow in serial but with exceptional scalability.

In order to extract a measure of absolute timing from our tests, we need to be able to effectively model system size scaling. For a conventional DFT code that calculates the eigenvalues and eigenvectors of the Kohn-Sham equation~\cite{ks}, either by explicit diagonalization (as we do here) or by an iterative minimization algorithm, it is well known that the calculation time scales cubically with system size (i.e., the number of atoms/molecules/basis orbitals). Linear-scaling methods~\cite{Bowler2012}, which make use of approximate spatial truncations based on the principle of electronic nearsightedness~\cite{near}, are also now well established and have been implemented in a number of popular codes.

The bottom panel of Fig.~\ref{fig:alpha} shows the results of all timing tests performed on two machines, JUQUEEN and SuperMUC; we plot the execution time for all number of cores, extrapolated to that of a single core as $t_{N_c} \mathbb{S}_1$ (from Eq.~\ref{eq:law1}), against the system size (the number of water molecules $N_m$). The estimated speedup $\mathbb{S}_1$ is obtained for any value of $N_m$ by using the fits of $S \left ( N_m \right )$ to the data in the top panel of Fig.~\ref{fig:law}, as described in the previous section. The resulting plot very clearly shows an almost pure cubic scaling with system size for both machines (the linear fits on the log--log scale have a fixed slope of 3). There is an excellent agreement in the extrapolated timings for each system size independently of the number of cores, and, even more encouragingly, our estimate of $\mathbb{S}_1$ appears to be robust even when extrapolating beyond the range of $N_m$ used in the fitting of $S$.

From these results, we can justify the use of a basic single-parameter model for system size scaling, of the form $t_1 = \alpha N_m^3$; lower-order terms are negligible even for the smallest system sizes considered here; this is because all the default routines in SIESTA other than the diagonalization procedure itself are linear-scaling by design. Within an SCF iteration, the contribution from building the sparse Hamiltonian matrix only become comparable to diagonalization for very high values of the cutoff energy defining the real-space grid, or non-local xc functionals such as those including dispersion interactions. We note here that we have also analyzed the strong scaling of individual SIESTA modules, finding diagonalization to be the bottleneck within an SCF iteration, while Hamiltonian construction is very efficient when using the parallelization strategy for the grid operations of Sanz-Navarro {\em et al.}~\cite{Sanz-Navarro2010} (accessible via the flag {\tt -DBSC\_CELLXC} at compilation).

The parameter $\alpha$, obtained by the fits shown in the bottom panel of Fig.~\ref{fig:alpha}, can therefore be used to compare the speed of the various machines, independently of differences in scaling performance. The values of $\alpha$ obtained for all six machines are shown in the top panel of Fig.~\ref{fig:alpha}. The large variation in $\alpha$ over almost two orders of magnitude is a reflection not simply of the machines' processor speeds (listed in Table~\ref{table:specs}), but also of numerous other interacting factors, such as the efficiency of the different compilers and libraries. In general, torus machines, which exhibit the best scaling, are predicted to be the slowest in serial, while fat tree machines, which do not scale as well in parallel, are predicted to be the fastest.

\begin{figure}
\begin{center}
\includegraphics{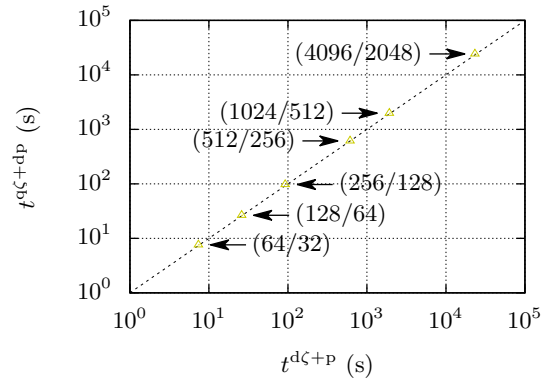}
\end{center}
\caption{Timing comparison on Curie for two different SIESTA basis sets. Each data point plots the execution time of a particular system size simulated with the \DZP\ basis (23 NAOs/H$_2$O molecule) against that of a {\em different} system size simulated with the \QZDP\ basis (46 NAOs/H$_2$O molecule), chosen so that the two systems have the same total number of basis orbitals. The two system sizes are shown in brackets (\DZP/\QZDP); in each case, both simulations are performed on the same number of cores, equal to the number of molecules in the \QZDP\ system.}
\label{fig:basis}
\end{figure}

We can now calculate a rough estimate of the execution time on each machine for any number of water molecules on any number of cores, by using our fits of the function $S \left ( N_m \right )$ and the parameter $\alpha$, and, hence, build up a $N_m$--$N_c$ `phase diagram' of the machine with the lowest execution time. This is shown in Fig.~\ref{fig:phase}: the main panel compares real timings, while the inset uses the estimates based on our fits. The agreement is best for large system sizes and number of cores, with some discrepancies appearing for $N_m \le 256$; this is not surprising, due both to the extrapolation of $S \left ( N_m \right )$ to low values, and the fact that the timings are very close for more than one machine.

The machines which gives the lowest absolute timings over the entire tested range of $N_c$ are overwhelmingly those with fat tree topologies, despite their inferior strong scaling performance with respect to torus machines. Two large regions can be clearly identified: Curie (BullX architecture) is the fastest machine for simulations with $N_c \lesssim 128$, while SuperMUC (IBM iDataPlex architecture) is the fastest above this value. There is some indication, confirmed by the model, that for large system sizes ($N_m \gtrsim 4096$) MareNostrum (IBM iDataPlex architecture) becomes faster than both of these machines (this might seem surprising, since it has the lowest value of $\alpha$, and, hence, should be the fastest in serial at all system sizes; however, it also exhibits the worst parallel scaling, making it less efficient than other machines for parallel calculations on even very few cores at modest system sizes). It is only for extremely large $N_c$ that the qualitatively different decrease in $S \left ( N_m \right )$ of the torus machines is predicted to lead to the lowest absolute timings, in particular for Hermit (Cray XE6 architecture), as it has a significantly lower $\alpha$ value than the IBM BlueGene/Q machines.

Our fitted models for the six supercomputers can also be used in a broader context, to estimate the execution time for any typical SIESTA calculation on HPC systems similar to the ones tested here. In fact, since the timing is dominated by the diagonalization procedure, especially for large system sizes, we can base our estimation on only two parameters, the total number of basis orbitals and the number of SCF iterations; we can safely neglect, to a first approximation, other parameters such as the number of electrons and the number and type of ions. This is illustrated in Fig.~\ref{fig:basis}, in which we compare calculations using the standard \DZP\ basis to ones using a larger \QZDP\ basis~\cite{basis_us} (twice the number of NAOs per water molecule) over a wide range of number of cores; as can be seen, timings on a given number of cores depend only on the total number of basis orbitals, and so a calculation using the larger basis takes the same time as one using the smaller basis with twice the system size. We note that this simple behaviour is due to the use of a solver which computes all eigenvalues by explicit diagonalization. Instead, solvers based on iterative minimization techniques (typically employed by plane-wave codes) scale only quadratically with the number of basis functions~\cite{OMM}; for such codes, we would expect the dependence of $S \left ( N_m \right )$ on basis size to be non-trivial. Unfortunately, we are not aware of published data for any other DFT code that could help in investigating this issue.

In order to allow SIESTA users to obtain absolute timing estimates for their parallel calculations, we have released a web applet~\cite{siestimator} based on the model we have described and the quantitative data obtained from our scaling tests. We also include a version of the applet for offline use in the Supporting Information ({\tt Code\_S1.txt}); details of the fits and the final set of parameters for the six machines can be easily found within the code.

\subsection{Weak scaling}

Finally, we briefly discuss the weak scaling behaviour demonstrated by the code. Weak scaling is of most interest to linear-scaling DFT codes~\cite{Bowler2010,VandeVondele2012}, for which the objective is to obtain a constant time-to-solution as the problem size is increased together with the number of cores (this is also known as Gustafson's law~\cite{Gustafson1988}). Cubic-scaling codes, instead, can achieve at best a quadratic weak scaling behaviour, which is rarely investigated~\cite{GPAW-GPU,OMM}; nevertheless, it can provide useful information on the limits of efficiency of the code.

\begin{figure}
\begin{center}
\includegraphics{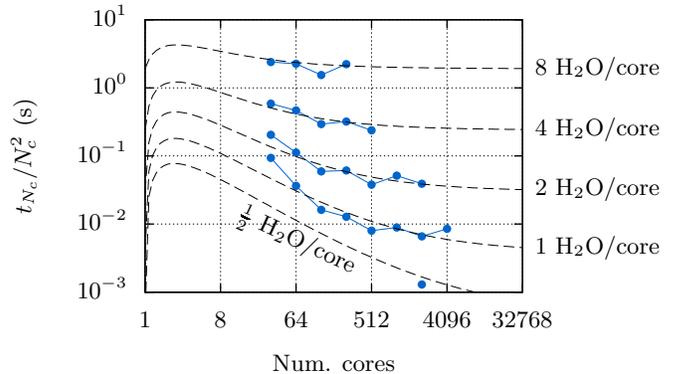}
\end{center}
\caption{Weak scaling on JUQUEEN for different numbers of water molecules per core. The execution time is divided by the square of the number of cores. The dashed lines show the estimates given by the fits of $S \left ( N_m \right )$ and $\alpha$.}
\label{fig:soft}
\end{figure}

We find it convenient to plot the execution time divided by the square of the number of cores, so that ideal weak scaling behaviour will appear flat, analogously to the case of a linear-scaling code. A representative example for one machine, JUQUEEN, is shown in Fig.~\ref{fig:soft}. Surprisingly, we observe better than ideal weak scaling, tending towards ideal as the number of cores is increased. The effect becomes more pronounced as the number of water molecules per core is decreased. These trends are almost perfectly reproduced by the timing estimates provided by our combined modelling of strong scaling and system size scaling.

We can understand this behaviour as a change in efficiency (as defined in the bottom panel of Fig.~\ref{fig:law}) due to the interplay between the decrease of $S$ with $N_m$ and the increase of $N_c$. Similarly, it is interesting to note that system size scaling at a fixed number of cores $>1$ deviates from its ideal cubic behaviour in serial.

If we assume $S \left ( N_m \right )$ to be of the form $A N_m^{-B}$, it is easily verified from the model that the weak scaling behaviour will tend towards ideal for $B \ge 1$; in the case of JUQUEEN, the fit gives a value of 1.8. This result applies equally to linear- and cubic-scaling codes, when using the appropriate definition of ideal weak scaling for each; indeed, a strikingly similar behaviour is reported for the linear-scaling Conquest code~\cite{Bowler2010}. As noted previously, the three machines with fat tree topologies appear to exhibit a slowing down in the decrease of $S$ with system size; although this should eventually make the weak scaling less than ideal, in practice it is not noticeable within the range of cores considered.

\section{Conclusions}

In this paper, we have investigated the performance of the SIESTA code on the six supercomputers of the PRACE Tier-0 network, currently amongst the largest in Europe. We propose a systematic investigation of parallel scaling using self-consistent calculations of snapshots of liquid water, varying both the number of cores on which the simulation is run and the number of water molecules per core; the largest simulation performed in our tests is of 8192 molecules on 8192 cores.

The results are analyzed using Amdahl's law to fit the data for each system size, providing a quantitative estimate of the code's efficiency over all number of cores based on a single parameter $S$; the scaling performance of the code, therefore, is completely described by the curve of $S$ as a function of system size. We find a qualitative difference in this curve depending on the topology of the connections between nodes in the supercomputer, with machines implementing torus topologies demonstrating a better scalability to large system sizes than those implementing fat tree topologies. Despite this, however, the latter group is shown to give lower absolute timings for almost all simulations within the tested range, as the performance on individual cores is significantly faster; furthermore, such architectures tend to offer a larger amount of memory per core, which can become an important issue either when running on few cores, or as the size of the simulation is increased (the memory requirements scale approximately quadratically with system size).

Combining Amdahl's law for strong scaling with a basic one-parameter model for system size scaling, both of which are fitted to the data provided by our tests, we can calculate a simple estimate of the execution time on a given number of cores for a generic total energy calculation with SIESTA; a new web applet~\cite{siestimator} developed in conjunction with the paper allows users of the code to employ this model for planning their projects on parallel architectures. An estimate of the memory requirement per core is also included.

Throughout the paper we have emphasized potential points of comparison with other DFT and electronic structure codes. Investigating and reporting $S \left ( N_m \right )$ curves for different HPC systems could provide valuable information to practitioners in the field, as well as for the ongoing development of the codes themselves. Care must be taken, however, when interpreting the results of comparisons based on strong scaling data, due to the fundamental differences between codes. Basis sets offer perhaps the most important example: is it meaningful to compare the strong scaling performance of a localized-orbital code and a plane-wave code for the same physical system? It is clear that $S$ varies with basis size, and so is crucially dependent in both cases on the precision level of the calculation; even disregarding the technical challenges involved~\cite{basis_us}, attempting to equate the two bases is not necessarily appropriate, as the codes are designed from the outset to be used with different aims. For this reason, we suggest that the best approach should not be overly competitive; rather, the objective should be to report on calculations using the typical setup appropriate for each code (e.g., the default \DZP\ basis for SIESTA), or possibly a range of different setups, as this will provide the most useful information for its users.

\begin{acknowledgments}
We thank Emilio Artacho, Alberto Garcia, and Georg Huhs for useful discussions. We acknowledge PRACE for awarding us access to the following resources: Hermit based in Germany at the High Performance Computing Center Stuttgart (HLRS), JUQUEEN based in Germany at the J\"{u}lich Supercomputing Centre, FERMI based in Italy at the CINECA SuperComputing Application and Innovation Department, Curie based in France at the Tr\`{e}s Grand Centre de Calcul du CEA (TGCC), SuperMUC based in Germany at the Leibniz Supercomputing Centre, and MareNostrum based in Spain at the Barcelona Supercomputing Center.
\end{acknowledgments}

\end{document}